\documentclass[aip,jap,reprint,a4paper,showpacs,showkeys,amsmath,amssymb,nobibnotes,altaffilletter,superscriptaddress]{revtex4-1}

\usepackage[utf8x]{inputenc}
\usepackage{graphicx}
\usepackage{pslatex}
\usepackage{xspace}
\usepackage{textcomp}	

\newcommand{\pht}{poly(3-hexyl thiophene)\xspace}
\newcommand{\pcbm}{[6,6]-phenyl-C$_{61}$ butyric acid methyl ester\xspace}
\newcommand{\degree}{$^{\circ}$}

\newcommand{\mytilde}{\raise.17ex\hbox{$\scriptstyle\mathtt{\sim}$}}

\graphicspath{{./}{images/}}

\begin{document}


\title{Nongeminate recombination in neat P3HT and P3HT:PCBM blend films}

\author{Julien Gorenflot}%
\author{Michael C. Heiber}%
\affiliation{Experimental Physics VI, Julius-Maximilians-University of W{\"u}rzburg, 97074 W{\"u}rzburg, Germany}

\author{Andreas Baumann}
\affiliation{Experimental Physics VI, Julius-Maximilians-University of W{\"u}rzburg, 97074 W{\"u}rzburg, Germany}%
\affiliation{Bavarian Centre for Applied Energy Research (ZAE Bayern), 97074 W{\"u}rzburg, Germany}

\author{Jens Lorrmann}%
\author{Matthias Gunz}%
\author{Andreas K\"ampgen}
\affiliation{Experimental Physics VI, Julius-Maximilians-University of W{\"u}rzburg, 97074 W{\"u}rzburg, Germany}

\author{Vladimir Dyakonov}\email{dyakonov@physik.uni-wuerzburg.de}
\affiliation{Experimental Physics VI, Julius-Maximilians-University of W{\"u}rzburg, 97074 W{\"u}rzburg, Germany}%
\affiliation{Bavarian Centre for Applied Energy Research (ZAE Bayern), 97074 W{\"u}rzburg, Germany}

\author{Carsten Deibel}\email{deibel@disorderedmatter.eu}%
\affiliation{Experimental Physics VI, Julius-Maximilians-University of W{\"u}rzburg, 97074 W{\"u}rzburg, Germany}

\date{\today}

\begin{abstract}
The slow decay of charge carriers in polymer--fullerene blends measured in transient studies has raised a number of questions about the mechanisms of nongeminate recombination in these systems.  In an attempt to understand this behavior, we have applied a combination of steady-state and transient photoinduced absorption measurements to compare nongeminate recombination behavior in films of neat \pht (P3HT) and P3HT blended with \pcbm (PCBM).  Transient measurements show that carrier recombination in the neat P3HT film exhibits second-order decay with a recombination rate coefficient that is similar to that predicted by Langevin theory.  In addition, temperature dependent measurements indicate that neat films exhibit recombination behavior consistent with the Gaussian disorder model.  In contrast, the P3HT:PCBM blend films are characterized by a strongly reduced recombination rate and an apparent recombination order greater than two.  We then assess a number of previously proposed explanations for this behavior, including phase separation, carrier concentration dependent mobility, non-encounter limited recombination, and interfacial states.  In the end, we propose a model in which pure domains with a Gaussian density of states are separated by a mixed phase with an exponential density of states.  We find that such a model can explain both the reduced magnitude of the recombination rate and the high order recombination kinetics and, based on the current state of knowledge, is the most consistent with experimental observations.
\end{abstract}



\maketitle

\section{Introduction}

Organic solar cells based on polymer--fullerene blends have recently reached power conversion efficiencies as high as 10\%.~\cite{green2012review} Paradoxically, those high performances are achieved due to an only marginally understood peculiarity: the inefficiency of nongeminate charge carrier recombination.  Such inefficient recombination has been observed in a number of polymer--fullerene blends.\cite{pivrikas2005a,parkinson2008,clarke2011,zusan2014}

The nongeminate recombination of two oppositely charged particles is predicted by the Langevin theory.\cite{langevin1903} This theory states that the recombination rate of electrons and holes ($R_\text{L}$) is governed by a second-order process,
	\begin{equation}
		R_\text{L} = k_\text{L}np.
		\label{eqn:bimol}
	\end{equation}
where $n$ is the electron concentration, $p$ is the hole concentration, and $k_\text{L}$ is the Langevin rate constant.  This rate constant is defined by assuming that the recombination event is much faster than the rate at which electrons and holes encounter one another.  As a result, the Langevin rate constant depends on the charge carrier mobility of each species, 
	\begin{equation}
		k_{\text{L}} = \frac{e}{\varepsilon} (\mu_e + \mu_h) .
		\label{eqn:prefact}
	\end{equation}
where $e$ is the elementary charge, $\epsilon$ the dielectric constant, $\mu_e$ is the electron mobility, and $\mu_h$ is the hole mobility. 

This implies that for equal densities of electrons and holes, $R \propto n^2$, and an initial carrier density ($n_0$) will decay as $n(t) = n_0/(1+k_\text{L}n_0t)\propto t^{-1}$ in the absence of further photogeneration. For conditions typical of an organic solar cell, with an initial polaron density of 10$^{18}$~cm$^{-3}$, a mobility of 10$^{-4}$~cm$^2$/Vs, and a relative dielectric constant of 3.5, Langevin theory predicts that almost 85$\%$ of charge carriers should recombine nongeminately within 100~ns, which is the minimum time required for the charge extraction.\cite{guo2010a,deibel2010review} Yet, external quantum efficiencies of 80$\%$ and higher have been reported under appropriate illumination in polymer--fullerene blends.~\cite{deibel2010review} 

In \pht (P3HT):\pcbm (PCBM) blends, the actual recombination rate is found to be up to 10$^{4}$ times slower than predicted by the Langevin theory,\cite{pivrikas2005a,deibel2009} resulting in the characterization of a reduction factor ($\zeta$),\cite{deibel2008b}
\begin{equation}
\zeta = \frac{R_\text{exp}}{R_\text{L}}.
\label{eqn:zeta}
\end{equation}
 Furthermore, a variety of experimental methods have shown that the recombination rate in polymer--fullerene blends does not have the expected second-order kinetics. Instead, higher orders between 2.3 and 2.8 have been found at room temperature.\cite{nelson2003,guo2010a,foertig2009,shuttle2010} Although under these conditions it is not possible to describe the higher order decay by reduced Langevin recombination based on Eqn.~(\ref{eqn:bimol}), such reduction factors are still being reported.\cite{kniepert2011,mingebach2012,wetzelaer2013} This discrepancy needs to be resolved in order to understand the detailed processes involved in nongeminate recombination.

In this paper, we compare the dynamics of nongeminate recombination in a polymer--fullerene blend (P3HT:PCBM) to those in the neat polymer (P3HT) in order to better understand the origins of the reduced recombination rate and super-second order kinetics. Using pump-probe transient absorption spectroscopy (TA), we measure the polaron decay dynamics from the 10 ns to the 100 {\textmu}s timescale from 59-300~K. In spite of numerous studies concerning charge generation in neat P3HT and P3HT:PCBM blends, as well as charge recombination in P3HT:PCBM blends,\cite{piris2009,guo2010a,grzegorczyk2010,howard2010,ohkita2008} TA studies of the nongeminate recombination dynamics in neat P3HT are so far missing. Based on these results, we discuss the feasibility of several proposed models for nongeminate recombination in polymer--fullerene blends.  

\section{Experimental Methods}

The experimental setup for steady state photo-induced absorption (PIA), as well as sample preparation, have been described elsewhere.~\cite{delgado2009} P3HT was purchased from BASF (Sepiolid P200) and PCBM from Solenne. All materials were used without further purification.  Solar cells prepared with these batches typically reach efficiencies over 3$\%$.~\cite{schafferhans2010}

All materials were dissolved in chlorobenzene at a concentration of 20 mg/ml. The films were deposited onto sapphire substrates by spin-coating and annealed at 140{\degree}C for 10 min. Blends with a 1:1 weight ratio were studied. Films were prepared under a nitrogen atmosphere in a glove box.  The thickness of the TA films was measured at approximately 300~nm by a profilometer. For TA experiments, samples were excited by a 5~ns pulse of a nitrogen/dye laser at a wavelength of 500~nm with a pulse energy of 25~{\textmu}J/cm$^2$. The generated polarons were probed using their characteristic absorption at 980~nm by an 80~mW cw laser. The decay of this absorption was measured using a FEMTO HCA-S-400M-IN preamplified InGaAs photodiode and recorded by a Tektronix oscilloscope.\cite{deibel2010review,guo2010a} The change in optical density ($\Delta OD$) was computed from the transient signal and is directly related to the density of the absorbing species by the absorption cross section.\cite{kroeze2004}

The hole mobility of neat P3HT films was also measured using the charge carrier extraction technique, OTRACE, as described in more detail elsewhere.\cite{baumann2012} OTRACE samples were prepared by spin-coating a solution of P3HT dissolved in chlorobenzene (30~mg/ml), resulting in 200 nm films as measured by a profilometer. For OTRACE measurements, the sample was directly transfered to a closed-cycle He cryostat without any exposure to air.  A pulsed 10 W neutral white Rebel-LED was then used to generate charge carriers in the bulk of the P3HT film.  The waveform was applied by an Agilent A81150A waveform generator, and the current transients were amplified by a FEMTO DHPCA-100 current amplifier and then recorded using an Agilent DSO90254A digital storage oscilloscope.
    
\section{Experimental Results}

\subsection{Steady state photoinduced absorption}

\begin{figure}[h]
    \includegraphics[width=8cm]{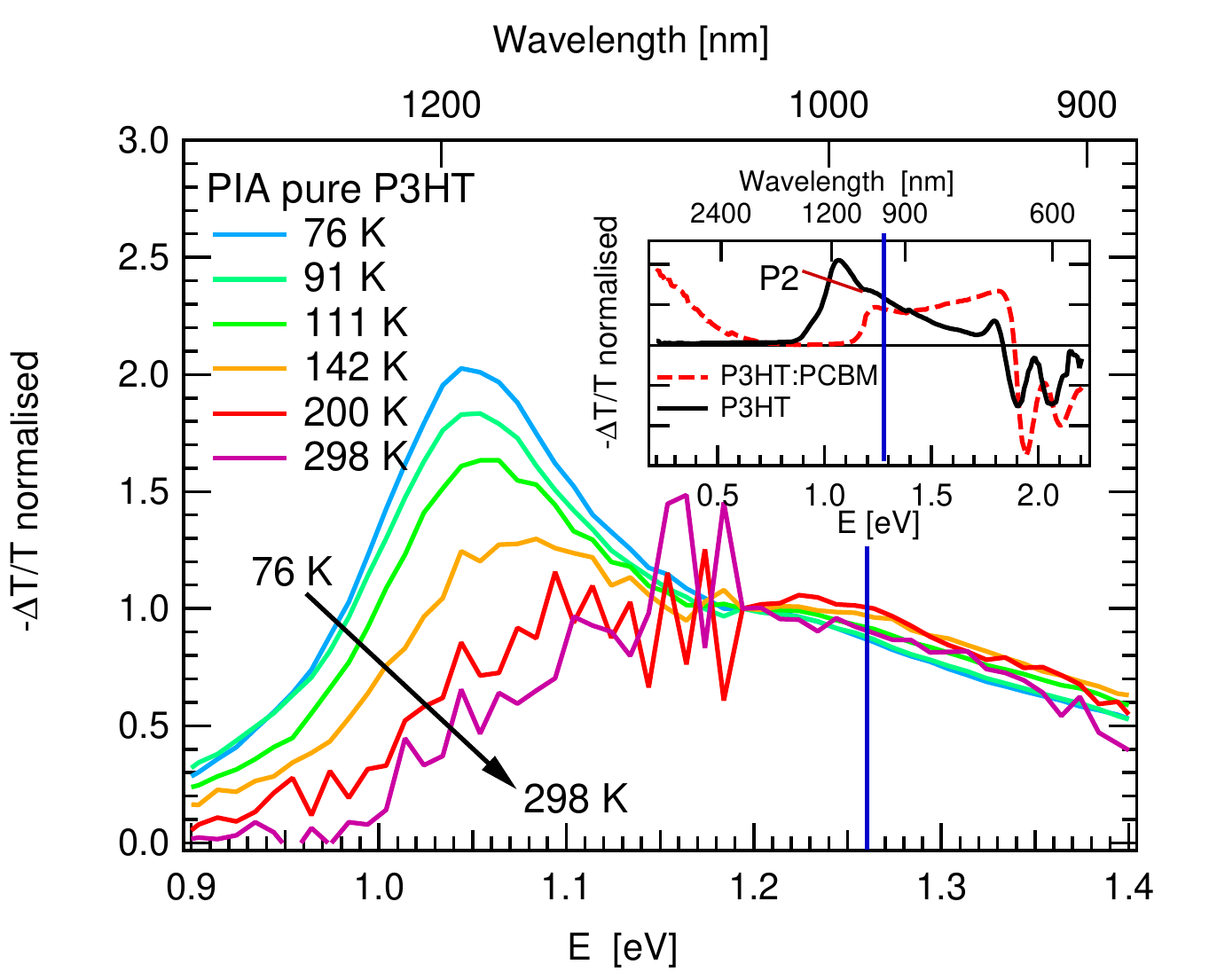}%
    \caption{(Color online) Steady-state PIA spectra of neat P3HT (normalized to absorption at 1.2~eV). The blue line indicates the probe wavelength used for TA; above 140~K, absorption at this wavelength is proportional to the density of polarons in both P3HT and P3HT:PCBM. Inset: Steady-state PIA spectra of neat P3HT (continuous line) and P3HT:PCBM blend (broken line) at a temperature of 30~K.%
    \label{fig:PIA}}
\end{figure}

In neat P3HT films, several species can coexist due to the lower efficiency of charge carrier photogeneration.\cite{deibel2010} At low temperatures, PIA spectra indeed exhibits bands due to several species as shown in Fig.~\ref{fig:PIA}. In addition to the polaronic features visible in the P3HT:PCBM blend films, there is a peak at 1170~nm (1.06~eV) that has been attributed to neutral species in P3HT.\cite{jiang2002a,vanhal1999} At 30~K, the tail of this peak is overlapping with the \textit{P2} polaron peak (see Fig.~\ref{fig:PIA} inset). Yet, at room temperature, the lifetime of those excitonic species becomes too short to contribute to PIA after 1~ns.\cite{piris2009} We find that their contribution to the absorption at 980~nm continuously decreases when increasing the temperature over 30~K and vanishes above 142~K (Fig.~\ref{fig:PIA}). 

In contrast, the density of the neutral species in P3HT:PCBM blends decays on the sub-nanosecond scale even at low temperatures.\cite{grzegorczyk2010} This decay is explained by efficient charge carrier photogeneration as revealed by the high external quantum efficiency measurements in solar cells based on this blend.\cite{brabec2001} It is therefore safe to assume that polarons are the only absorbing species at 980~nm in the blend over the time range $10^{-8}$ to $10^{-4}$~s. We conclude that, above 140~K, the absorption at 980~nm is representative of the polaron density in both the neat and blend films.

\subsection{Transient absorption: absorption cross section and photogeneration}

TA was probed at 980~nm, which corresponds to the maximum of the so-called \textit{P2} peak (see inset of Fig.~\ref{fig:PIA}). A number of spectroscopic studies including spin-sensitive methods have assigned this peak to P3HT polaron absorption.\cite{jiang2002a,brown2001} The corresponding absorption cross section was determined as follows. For a P3HT:PCBM blend film excited by a  25~{\textmu}J/cm$^2$ pulse of 500~nm light, 10.3$\%$ of the incident light was reflected by the sample and 3.6$\%$ was transmitted, yielding an upper limit for the generated exciton density of 1.8$\times 10^{18}$~cm$^{-3}$. By solving the rate equations described by \citeauthor{howard2010},\cite{howard2010} including the nonlinear losses due to polaron--exciton annihilation at high excitation intensity,~\cite{ferguson2011} and implementing a Gaussian exciton generation term to represent the laser pulse, we estimated the number of generated positive polarons to be (7.8$\pm$2.0)$\times 10^{17}$~cm$^{-3}$.  
We also obtained a polaron--exciton annihilation rate of (1.7$\pm$0.7)$\times 10^{-7}$~cm$^3$/s from the excitation intensity dependence of the initial change of optical density ($\Delta OD_0$). From the corresponding change in the optical density ($\Delta OD$), the absorption cross section of P3HT polarons at 980~nm in blend films was evaluated at (1.9$\pm$0.5)$\times 10^{-16}$~cm$^2$, which is in a similar range as recently reported for neat P3HT films.\cite{leijtens2013}

\begin{figure}
    \includegraphics[width=8cm]{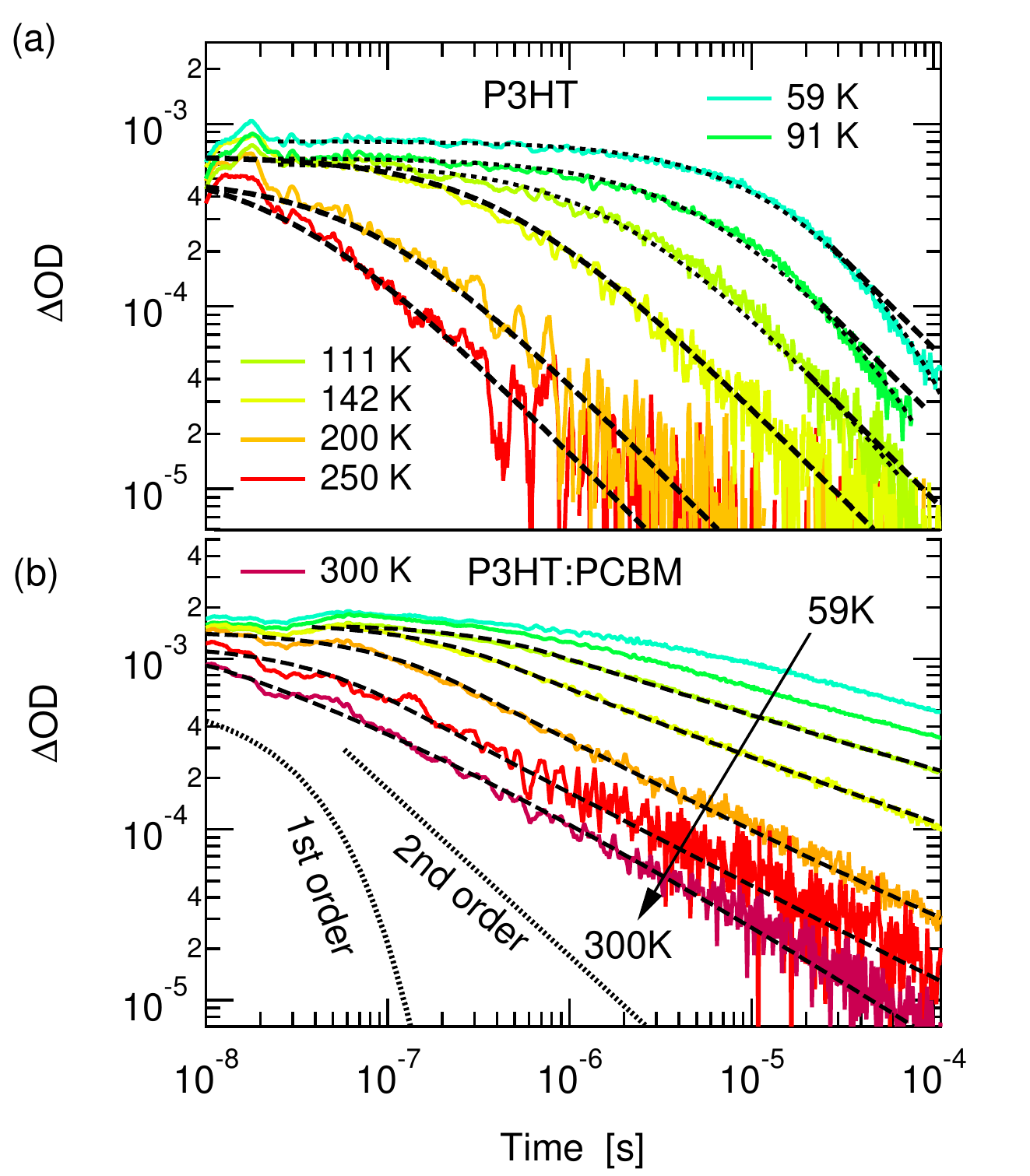}%
    \caption{(Color online) (a) Transient absorption decays in neat P3HT for different temperatures (solid lines). The dashed lines show fits including only second-order recombination and the dotted lines show fits accounting for contributions by both first and second-order decay. The asymptotes corresponding to a purely second-order decay are shown for comparison for 111~K, 91~K and 59~K (dashed lines). (b) Transient absorption decays in P3HT:PCBM (solid lines) and power law fits (dashed lines). Typical shape of first and second-order decays (dotted lines) are shown for comparison.
    \label{fig:transients}}
\end{figure}

The transient decays of the change in optical density for the neat P3HT and P3HT:PCBM blend films are shown in Fig.~\ref{fig:transients}. The initial change of optical density at 10~ns  ($\Delta OD_0$), which includes geminate recombination and exciton--polaron annihilation,~\cite{howard2010} is virtually temperature independent and is only slightly lowered at temperatures approaching 300~K due to a faster onset of nongeminate losses at higher temperatures. This finding is consistent with earlier reports of temperature independent charge carrier photogeneration in polymer--fullerene blends.\cite{pensack2009,grzegorczyk2010,street2010a} For both neat and blend films, the polaron decay beyond 10~ns is due to nongeminate recombination, with an increasing recombination rate for higher temperatures. In the following sections, we will focus on these nongeminate losses.

\subsection{Transient absorption: neat P3HT} \label{sec:exp-p3ht}

In order to gain a deeper understanding of the polaron dynamics, we compare the experimental decays to analytical models based on continuity equations. In the absence of any external contributions (injection or photoexcitation from the ground state) after exciton generation by the laser pulse at $t=0$, the continuity equation describing the total polaron density ($n$) is
    \begin{equation}
        \frac{dn}{dt} = -R
        \label{eqn:continuity}
    \end{equation}
for $t>0$, where $R$ is the recombination rate.

Although Langevin recombination (Eqn.~(\ref{eqn:bimol})) is the expected loss mechanism for separated polarons, we also consider a first order decay.\cite{deibel2010review} The sum of a first and a second-order term is able to perfectly fit the decays observed in neat P3HT at temperatures below 140~K, whereas at higher temperatures, the decays are found to be purely second-order.  These findings are in agreement with the assignment of the absorption signal in neat P3HT at 980~nm to polarons at temperatures above 140~K and to a sum of contributions from both polarons and neutral species at lower temperatures. These neutral species could be triplet excitons,\cite{vanhal1999} interchain singlet excitons,\cite{jiang2002a} or polaron pairs~\cite{howard2010} and are outside the scope of this article. 
    
\begin{figure}[t]
	\includegraphics{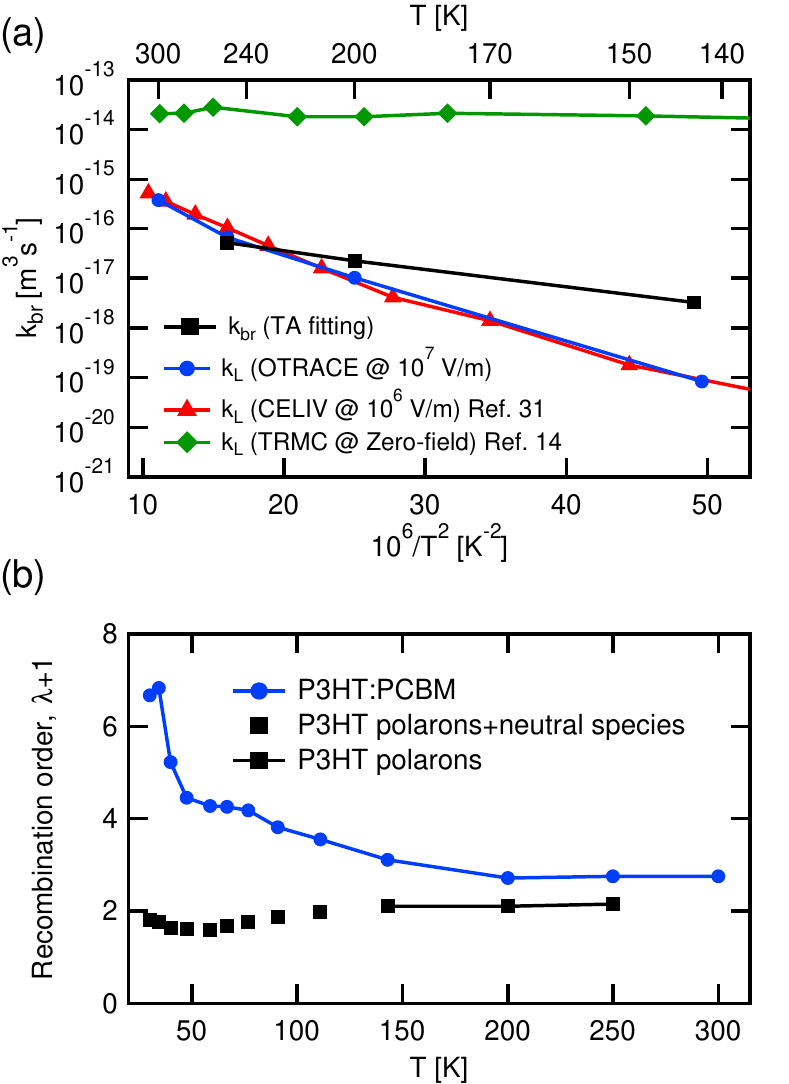}
	\caption{(Color online) (a) Recombination coefficients extracted from neat P3HT transients (Fig.~\ref{fig:transients}) compared to Langevin coefficients calculated from experimental mobility measurements using Eqn.~(\ref{eqn:prefact}) (b) Apparent recombination order as function of temperature for neat P3HT and P3HT:PCBM blend films.}
	\label{fig:rec_order}
\end{figure}

As predicted by Langevin theory, the recombination of polarons in neat P3HT exhibits second-order kinetics as shown in Fig.~\ref{fig:rec_order}b. In addition, the temperature dependence of the second-order recombination coefficient ($k_\text{br}$) obtained from the fits in Fig.~\ref{fig:transients}a is compared to Langevin recombination coefficients ($k_\text{L}$) calculated from temperature dependent mobility measurements in Fig.~\ref{fig:rec_order}a. Assuming equal electron and hole mobilities and a dielectric constant of 3.5, Eqn.~(\ref{eqn:prefact}) was used to calculate the Langevin recombination rate coefficient from several different mobility measurements.  The recombination coefficient derived from the transients is very similar to Langevin theory at 250~K when compared to coefficients derived from our OTRACE experiments and from previous CELIV measurements.\cite{mozer2005a} ToF measurements also indicate similar magnitudes and temperature dependencies as the OTRACE and CELIV measurements shown here.\cite{mozer2005a,mauer2010}  However, the rate coefficients determined from the transients demonstrate a much weaker temperature dependence than observed in mobility measurements. At 150~K the measured recombination coefficient is more than one order of magnitude greater than expected from Langevin theory, although it remains far below the calculated Langevin recombination rate using the temperature independent local mobility determined by time-resolved microwave conductivity (TRMC).\cite{grzegorczyk2010}

\subsection{Transient absorption: P3HT:PCBM blend}

While the dynamics of charge recombination in neat P3HT appear Langevin-like, the transient absorption signal in the P3HT:PCBM blend exhibits a much slower decay (Fig.~\ref{fig:transients}b), which is similar to previous reports.~\cite{guo2010a,foertig2009,shuttle2010} It is characterized by a reduced recombination rate that does, however, depend on time (or carrier concentration). The recombination order exceeds 2 already at room temperature, increasing to about 7 at 30~K as shown in Fig.~\ref{fig:rec_order}b. Since the recombination mechanism is still assumed to be between one electron and one hole, the resulting experimentally determined nongeminate recombination rate is expressed as 
\begin{align}
	R_\text{exp} = k_{\text{exp}}(n) n^2 \propto  n^{\lambda+1},
	\label{eqn:R_exp}
\end{align}
where $k_\text{exp}$(n) is the carrier concentration dependent rate coefficient and $\lambda+1$ is the recombination order.  This form is convenient because the slope of the polaron decay on the log-log plots shown in Fig.~\ref{fig:transients} is equal to $-1/\lambda$.  To produce a recombination order of $\lambda+1$, the recombination rate coefficient must then take on the form
\begin{align}
	k_\text{exp}(n) \propto n^{\lambda-1}.
	\label{eqn:k_exp}
\end{align}

\section{Discussion}

\subsection{Multiple Trapping and Release (MTR) Model}

It is well known that charge carrier trapping plays a significant role in P3HT and other organic semiconductors.\cite{schafferhans2008} Therefore, we approach the interpretation of our experimental data using the multiple-trapping-and-release (MTR) model.\cite{schmidlin1977,noolandi1977,oelerich2012} With this model, the overall charge carrier density ($n$) is split into two populations, free carriers ($n_c$) and trapped carriers ($n_t$).  The free carriers are assumed to move at a speed defined by the free carrier mobility ($\mu_c$), and the trapped carriers are assumed to be immobile.  Over time, trapped carriers are thermally excited to become free carriers, and free carriers relax into trap states.  However, under steady state conditions, the ratio of free to trapped carriers is assumed to be constant.  To start our analysis, we assume that there is a very low concentration of intrinsic dark carriers, that the concentrations of electrons and holes are equal due to the symmetric nature of photogeneration, and that mobilities of electrons and holes are equal.  

\subsection{Neat P3HT: Langevin-like recombination}

To analyze the second-order decay observed in the neat P3HT measurements, it is then assumed that carrier recombination can only occur between free electrons and free holes, free electrons and trapped holes, or trapped electrons and free holes.\cite{kirchartz2011,foertig2012,deibel2013}  As a result, the recombination rate is given by
\begin{align}
	R \approx \frac{e}{\varepsilon} (2 \mu_{c} n_c^2 + 2 \mu_{c} n_c n_t) = \frac{e}{\varepsilon} 2 \mu_{c} n_c (n_c + n_t)	
	\label{eqn:R_neat}
\end{align}
We then define the fraction of free charge carriers, $\Theta$, where $\Theta=n_c/n$, resulting in a final recombination rate
\begin{align}
	R \approx 2\frac{e}{\varepsilon}\mu_c \Theta n^2.
	\label{eqn:R_theta}
\end{align}
As a result, the effective macroscopic mobility ($\mu$) is governed by the free carrier mobility and the fraction of free carriers, $\mu = \Theta\mu_c$.  Under these conditions, Eqn.~(\ref{eqn:R_theta}) is equivalent to Eqn.~(\ref{eqn:bimol}), and within the MTR model, $k_\text{L}$ can then be expressed as
\begin{equation}
k_\text{L} = 2 \frac{e}{\epsilon}\mu_c \Theta.
\label{eqn:MTRprefact}
\end{equation}  
Within this framework, we note that if $\Theta(n)$ is not constant, the macroscopic mobility would be expected to be carrier concentration dependent, resulting in super-second-order recombination.  However, if $\Theta(n)$ is constant, the macroscopic mobility would be independent of the charge carrier concentration, and second-order recombination is expected.  Considering the charge carrier dynamics measured here, we observe second-order decay between 140 and 300~K  (Fig.~\ref{fig:rec_order}), implying that neat P3HT can be well described by Eqn.~(\ref{eqn:bimol}) and that both the macroscopic mobility ($\mu$) and $\Theta$ are independent of the carrier concentration.  

We assume for now, in accordance with the findings of Oelerich et al.,\cite{oelerich2012} that P3HT has a Gaussian density of states (DOS) distribution.  Then, within the framework of the Gaussian disorder model (GDM), in which $\mu(T) \propto \exp( -( 2\sigma/3k_BT)^2)$,\cite{bassler1993} we determined the standard deviation of the DOS ($\sigma$) to be 37~meV. Previously, a value of 56~meV was found for holes by photocurrent transient measurements on much thicker samples (c.f. Fig.~\ref{fig:rec_order}a).\cite{mauer2010} Using the GDM, the macroscopic mobility in neat disordered materials has been considered previously by hopping master equation\cite{pasveer2005} and the MTR model.\cite{oelerich2012} For $\sigma=$37~meV and temperatures between 140 and 300~K, the mobility depends on the carrier concentration only if more than $10^{-3}$ of the states are occupied.\cite{pasveer2005} Below that limit, the mobility is predicted to be independent of the carrier concentration. In contrast, if neat P3HT has an exponential DOS, the mobility would be expected to depend on the carrier concentration in all regimes.\cite{oelerich2012} Therefore, the observed second-order decay implies a carrier concentration independent mobility that is consistent with a Gaussian DOS but not with an exponential DOS. This finding is consistent with the analysis of mobility measurements by Oelerich et al.\cite{oelerich2012}

However, as shown in Fig.~\ref{fig:rec_order}, we find that the measured recombination coefficient has a weaker temperature dependence than expected from OTRACE and CELIV hole mobility measurements, although not temperature independent as shown in the TRMC measurements.  TRMC measures the high frequency photoconductivity, which is assumed to result from the motion of free carriers and to be proportional to $n_c \mu_c$, in contrast to the macroscopic mobility ($\mu$) measured by the other techniques, which calculate the mobility based on long-range charge transport of all carriers.  As a result, mobility values derived from TRMC are usually much higher, but it is unclear why the temperature dependence is so weak.  We would expect that the free carrier concentration ($n_c$) should be temperature dependent and cause the TRMC signal to have a stronger temperature dependence.  This discrepancy makes it difficult to rely on mobility measurements derived from TRMC experiments when describing mechanisms that require longer range charge transport until further studies clarify the nature of the mobility measured by TRMC.  In any case, the experimentally determined rate coefficients are greater than the Langevin rates derived from macroscopic mobility measurements at lower temperatures and have a weaker temperature dependence.  

It is plausible that the mobility that characterizes the charge motion required for recombination is different than that for long range macroscopic charge transport.  For example, charge trapping may be effectively shallower for transport on the \mytilde 10~nm length scale than on the \mytilde 100~nm length scale due to spatial homogeneity arising from the presence of crystalline and amorphous domains.  Another possibility is that the electron and hole mobilities have different temperature dependencies.  However, further detailed studies are needed to test these concepts.  As a result, while Langevin theory works fairly well to describe nongeminate recombination near room temperature, questions remain as to why the temperature dependence is weaker than expected from macroscopic mobility measurements.  Nevertheless, the recombination rate has no major reduction factors and is actually slightly greater than expected at lower temperatures.  This implies that P3HT is sufficiently homogeneous such that mobile charge carriers can reach their recombination partners everywhere.

Comparing to previous recombination measurements on neat P3HT films, our observation of second-order recombination dynamics in neat P3HT films is in contrast to the first-order decays observed in TRMC measurements by Ferguson et al.\cite{ferguson2011}  In their study, they attribute the first-order decay to the presence of a significant dark carrier concentration (\mytilde $10^{19}$~cm$^{-3})$.  If the observed behavior is indeed due to the presence of dark carriers, our measurements suggest that the neat P3HT samples that we have tested have a significantly lower dark carrier concentration.   Our observation of second-order kinetics indicates that the dark carrier concentration in our samples is less than the range of photogenerated carrier concentrations tested. For an estimated initial exciton concentration of \mytilde$10^{18}$~cm$^{-3}$ due to the excitation laser pulse and an upper bound of \mytilde10\% carrier yield, we estimate the photogenerated carrier concentration tested here is in the range of \mytilde$10^{15}$~cm$^{-3}$ to \mytilde$10^{17}$~cm$^{-3}$.  As a result, we estimate that the dark carrier concentration in our samples is less than \mytilde$10^{15}$~cm$^{-3}$. 

\subsection{P3HT:PCBM blends: Langevin recombination?}

In contrast to neat P3HT, the P3HT:PCBM blend films clearly display slower recombination and super-second-order decay (Fig.~\ref{fig:rec_order}b).  Under these conditions, the reduction factor defined in Eqn.~\ref{eqn:zeta} and calculated using Eqn.~\ref{eqn:bimol} and Eqn.~\ref{eqn:R_exp} becomes carrier concentration dependent with the form 
\begin{equation}
\zeta = \frac{R_\text{exp}}{R_\text{L}} = \frac{k_\text{exp}(n)}{k_\text{L}} \propto n^{\lambda-1}.
\label{eqn:zeta2}
\end{equation}
To understand the origin of this reduction factor, we discuss several previously proposed hypotheses.  Based on the measurements presented here and in previous studies in the literature, we attempt to eliminate those that are inconsistent with the current state of experimental knowledge, highlight those that are still feasible, and finally direct researchers to areas where further measurements are needed.  We emphasize that any well-suited model must be able to account for both the magnitude and the carrier concentration dependence of the reduction factor.

Previously, it has been suggested that due to the presence of lamellar crystals in P3HT domains, which may promote two-dimensional transport, the resulting recombination behavior in P3HT:PCBM blends is more accurately represented by a two-dimensional Langevin recombination model.\cite{juska2009}  However, our observation that the neat P3HT films, which also have lamellar crystalline domains, do not demonstrate these same characteristics suggests that two-dimensional transport is not a dominant factor and that a two-dimensional Langevin recombination model is not appropriate.  In addition, kinetic Monte Carlo simulations implementing anisotropic mobility also conclude that the anisotropy effect is likely to be too weak to be the dominant factor in P3HT:PCBM blends.\cite{groves2008b}  Another previously suggested explanation for the super-second order recombination kinetics is the presence of carrier concentration gradients near the electrodes in operational devices.\cite{deibel2009}   While carrier concentration gradients may enhance this effect in devices, our observation here of similar kinetics on samples without electrodes suggests that is not likely be the dominant cause.  With these hypotheses ruled out, we move now to a more detailed discussion of the remaining concepts in the following subsections.

\subsubsection{Effect of phase separation} 

One important difference between traditional Langevin theory and the P3HT:PCBM blend system is the presence of a complex nanoscale phase separated morphology\cite{erb2005} that spatially limits the motion of the electrons and holes and limits the possible places where recombination can occur. The reduction factor has been previously attributed to the presence of phase separation.\cite{koster2006,baumann2011}  To assess the effect of phase separation, we derive and compare the expected recombination rate equations for a homogeneous blend and a phase separated blend with pure domains.  

For a homogeneous blend, similar to neat P3HT, in a system with charge carrier trapping due to energetic disorder,\cite{schafferhans2010,foertig2012} the recombination rate can be approximated by Eqn.~(\ref{eqn:R_theta}) and, equivalently, by Eqn.~(\ref{eqn:bimol}).  As a result, the standard Langevin rate equation would be expected in a homogeneous blend.  Moving now to a phase separated blend, carriers trapped in the interior of the domains are unable to undergo recombination, which should reduce the overall recombination rate.  If we assume that very few charge carriers are trapped close to the interface, the recombination rate should be dominated by reactions between free electrons and free holes.  In this framework, the resulting recombination rate is 
\begin{equation}
	R \approx 2\frac{e}{\varepsilon} \mu_{c} n_c^2 \approx \Theta k_\text{L} n^2.
	\label{eqn:R_ps}
\end{equation}
Here, the trapped charge carriers in the bulk of the domains are protected from recombination as long as they are trapped,\cite{baumann2011} lowering the overall recombination rate by a factor of $\Theta$. 

However, given the nanoscale dimensions of the domains ($\approx15$~nm),\cite{vanbavel2009,pfannmoeller2011} there can actually be a large fraction of the P3HT volume near the interface.  As a simple example, given a spherical domain with a 15~nm diameter, 35\% of the volume is within 1 nm of the interface, and it can be expected that 35\% of the trapped carriers can participate in recombination.  Therefore, the amount of carriers trapped close to interface is, in fact, not likely to be negligible.  By including recombination between free carriers and these carriers trapped near the interface, the recombination rate equation becomes
\begin{equation}
	R \approx \Theta k_\text{L} n^2 + \chi (1-\Theta) k_\text{L} n^2,
	\label{eqn:R_phi_i}
\end{equation}
where $\chi$ is the interfacial volume fraction, the fraction of the donor and acceptor volume at the interface with respect to the total volume. 

Furthermore, if we assume that most carriers at any given time are trapped ($\Theta\ll1$) and that $\Theta\ll\chi$, then
\begin{equation}
	R \approx \chi k_\text{L} n^2.
	\label{eqn:R_phi}
\end{equation}
As a result, this scenario predicts that the reduction factor is approximately equal to $\chi$.  However, the magnitude of $\chi$ expected in a nanostructured morphology ($>10^{-1}$) is closer to unity than previously measured reduction factors (\mytilde $10^{-3}$) at room temperature.\cite{deibel2008b}  In addition, a simple phase separation model has no way of explaining the super-second order kinetics.  As a result, we find it unlikely that phase separation inherently causes the apparent major deviations from Langevin theory.  

\subsubsection{Carrier concentration dependence of mobility}

Assuming that carrier recombination is still encounter-limited as assumed in Langevin theory, the recombination rate should still depend on the carrier mobility.  Up to now, we have assumed that the carrier mobility is independent of the carrier concentration, but a more complex carrier concentration dependence must be considered.  \citeauthor{shuttle2010} have attempted to explain the super-second order decay by assuming a carrier concentration dependent mobility in which $\mu(n) \propto n^{\lambda-1}$.\cite{shuttle2010} Such behavior would only be expected if the materials were to have an exponential DOS.  However, we have recently shown that this explanation may not generally hold,\cite{rauh2012} but we point out here that neither of these studies used a method that probes the charge carrier mobility directly. Therefore, these previous conclusions need to be verified.  In addition, \citeauthor{savenije2011} compared TRMC and TA measurements on P3HT:PCBM thin films and concluded that the mobility in P3HT:PCBM blends is time independent on the timescale of tens of nanoseconds onwards, indicating a carrier concentration independent mobility for the range tested.\cite{savenije2011}  However, as discussed in section IV.B on neat P3HT, TRMC mobility measurements may probe behavior that is significantly different than the more macroscopic mobility important for describing nongeminate recombination.  As a result, further concentration dependent mobility studies are needed to completely rule out carrier concentration dependent mobility as a main cause of the super-second order recombination kinetics.  

\begin{figure}
    \includegraphics[]{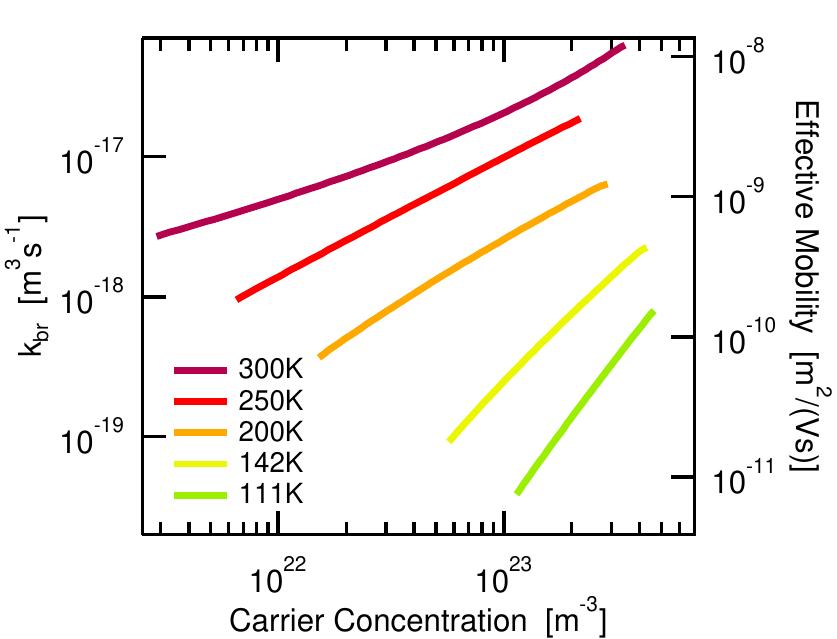}%
    \caption{(Color online) Predicted carrier concentration dependence of the mobility when assuming super-second order recombination is caused only by a carrier concentration dependent mobility.
    \label{fig:k_vs_n}}
\end{figure}

If a carrier concentration dependent mobility is to be the dominant cause of the observed recombination kinetics, this relationship should be proportional to the concentration dependence of the experimental recombination prefactor, $k_\text{exp}$, shown in Fig.~\ref{fig:k_vs_n}.  However, given that neat P3HT mobility is only weakly carrier concentration dependent, consistent with a Gaussian DOS, it is a reasonable assumption that at least the pure P3HT domains should demonstrate similar behavior.  But even if the P3HT domains do have a Gaussian DOS, another possibility is that the PCBM domains have an exponential DOS and a mobility that dominates the recombination rate.  However, space-charge limited current (SCLC) measurements on PCBM have indicated behavior consistent with a Gaussian DOS,\cite{mihailetchi2003} but SCLC measurements are also performed at much higher carrier concentrations than are present in working solar cells.  To clarify this behavior further, carrier concentration dependent mobility measurements on neat PCBM films are needed as well.  

\subsubsection{Effect of non-encounter limited recombination}

It has also been proposed that nongeminate recombination is not encounter limited as assumed in Langevin theory.\cite{hilczer2010,ferguson2011}  In this case, if the actual recombination mechanism itself is slow, the decay of charge carriers in the blend does not depend solely on the encounter probability.  Instead of the electron and hole recombining immediately when reaching each other, they form an intermediate polaron pair state that can recombine after some time but can also re-dissociate into free charges.  The experimental recombination rate is then defined
\begin{equation}
	R \approx -\frac{dn}{dt} = k_L n^2 - k_d [PP],
	\label{eqn:non-encounter}
\end{equation}
where $k_d$ is the polaron pair dissociation rate and $[PP]$ is the polaron pair concentration.  For this to have a major effect, the second term must be much larger than the Langevin term, which means the polaron concentration must persist over long timescales and the polaron pair dissociation rate must be fairly fast.  To determine the conditions in which this would occur, we first describe the polaron pair rate equation as
\begin{equation}
	\frac{d[PP]}{dt} = k_L n^2 - k_r [PP] - k_d [PP],
	\label{eqn:polaron-pair_rate}
\end{equation}
where $k_r$ is the rate of the final polaron pair recombination event by which the electron finally returns to the ground state.  For polaron pairs to persist in the system, $d[PP]/dt$ must not be much less than zero, and in a special case, when the polaron pair concentration is constant ($d[PP]/dt=0$) and the reduction factor ($\zeta$) is very small, 
\begin{equation}
	\zeta \approx \frac{k_r}{k_d} 
	\label{eqn:zeta_non-encounter}
\end{equation}
However, if the polaron pair concentration is changing over time, a more complicated expression is necessary.  Nonetheless, given the highly efficient charge separation that occurs in P3HT:PCBM blends, it is likely that the dissociation rate is significantly faster than the recombination rate.  As a result, while a significant reduction in the observed polaron decay rate could be expected, this model still does not explain the origins of the super-second-order kinetics.

\subsubsection{Effect of interfacial states}

Another important aspect to consider is the presence of a third mixed phase with unique materials properties compared to the pure phases.  P3HT:PCBM blends have been shown to have a more complex morphology than simply pure donor and pure acceptor domains,\cite{collins2010} and the presence of a mixed amorphous phase has been clearly identified.\cite{pfannmoeller2011}  In addition, the presence of deep states that go beyond a superposition of the tail states of the separate pure materials has been experimentally measured.\cite{vandewal2009a} These deeper states originate from the close interaction of donor and acceptor molecules at the heterointerface.  Taking this into account, Street et al.\ have proposed recombination via interfacial states.\cite{street2010,deibel2010b,street2010b}

With this in mind, we consider a model in which pure domains are separated by an interfacial mixed region that contains a DOS that is different from those present in either of the pure domains.  We note that implementing interfacial mixing without a separate DOS simply increases $\chi$, as previously defined in subsection 1, and cannot explain the observed behavior.  With separate DOS distributions, however, it is plausible that the macroscopic mobility would be dominated more so by charge transport within the pure domains containing a Gaussian DOS, as indicated by our measurements on neat P3HT, but that the actual recombination event is governed mainly by the spatial and energetic properties of the interfacial regions containing an exponential DOS, which has been indicated by defect spectroscopy.\cite{foertig2012,presselt2012}

To determine the expected recombination rate in this more complex scenario, we need to consider two separate contributions to the recombination rate, the behavior of the carriers in the pure domains and in the interfacial regions.  First, we assume that the majority of the carriers are trapped and that the contribution from free--free recombination is negligible.  As a result, the dominant recombination mechanism occurs when free carriers from the pure domains ($n_{c,p}$) are transported to the interfacial regions and recombine with carriers that are already present within the interfacial regions ($n_i$).  In this case, the resulting recombination rate becomes
\begin{equation}
	R \approx 2\frac{e}{\varepsilon}\mu_{c,p} n_{c,p} n_i,
	\label{eqn:R_interfacial}
\end{equation}
where $\mu_{c,p}$ is the mobility of the free carriers in the pure phases.  Rewriting this in terms of the overall carrier concentration ($n$), where $n=n_{c,p}+n_{t,p}+n_{c,i}+n_{t,i}$, the recombination rate becomes
\begin{equation}
	R \approx \Phi (1-\Phi) k_L n^2,
	\label{eqn:R_interfacial_n}
\end{equation}
where $\Phi$ is the fraction of carriers in the pure phase with respect to all carriers, $n_p/(n_p+n_i)$, and $k_L$ is derived from the mobility of the pure phases.  

This model can explain both the magnitude and the carrier concentration dependence of the reduction factor when $\Phi$ is large and carrier concentration dependent.  Given studies that have indicated an energetic driving force for carriers to diffuse from amorphous mixed regions to more ordered pure domains,\cite{mcmahon2011,jamieson2012} it is probable that $\Phi$ would be large, and given different DOS distributions, it is possible that the density of occupied states would be populated in different proportions at different overall carrier concentrations.  Here, to give super-second order kinetics, interfacial states would have to fill up proportionally faster than the states in the pure phases when increasing the overall carrier concentration.  Further theoretical and experimental studies are needed to test this model in more detail, but given the current state of knowledge and the critical analysis presented here, we find it to provide the most complete explanation to date.

\section{Conclusions}

To conclude, we have used transient absorption spectroscopy to monitor the nongeminate polaron decay in neat P3HT and P3HT:PCBM blend films. In the neat polymer, we observed Langevin-like recombination at temperatures above 140~K with second-order kinetics and a recombination coefficient that is slightly less temperature dependent than macroscopic mobility measurements. To analyze the results, we have used a multiple trapping and release (MTR) model to derive the expected recombination rate equations.  For neat materials, the MTR model predicts recombination dynamics consistent with Langevin theory, and the neat P3HT measurements appear to be mostly consistent with this model, aside from the weaker temperature dependence.  Most importantly, though, no significant reduction factor was observed in neat P3HT, and dark carriers were not found to play a role in the recombination kinetics.  

In contrast, the recombination dynamics in the blend films were characterized by a reduced recombination rate and super-second-order recombination kinetics. To narrow down the possible explanations for this behavior, we first eliminated several different models previously proposed, including a two-dimensional Langevin model and carrier concentration gradients.  To understand the effect of phase separation, the MTR model was used to derive the expected recombination rate for a homogeneous and phase separated blend.  However, phase separation alone was shown to only slightly reduce the recombination rate.  In addition, we argued that the mobility is not likely to be strongly carrier concentration dependent but have identified that further measurements are needed to rule it out as the sole contributor to the higher order recombination kinetics.  We then considered the idea that nongeminate recombination is not encounter limited as assumed by Langevin theory and found that the recombination rate could indeed be significantly reduced from the rate predicted by Langevin theory.  However this would still be unable to explain the origins of the recombination order.

Finally, we considered the effects of interfacial states, which have been previously identified and proposed to play a significant role in the recombination behavior.  Using the MTR model, we then derived the recombination rate expected when there are pure domains with a Gaussian density of states that are separated by a mixed interfacial phase with an exponential density of states.  This scenario is expected to produce both a reduced recombination rate and super-second-order recombination kinetics.  While still a qualitative model, we propose that it is most consistent with the available experimental data to date.  

\begin{acknowledgments}
The current work is supported by the Bundesministerium f{\"u}r Bildung und Forschung in the framework of the GREKOS project (contract no.~03SF0356B) and the European Commission through the Human Potential Program (Marie-Curie RTN SolarNType contract no.~MRTN-CT-2006-035533) and the Deutsche Forschungsgemeinschaft, DFG under the contract INST 93/623-1 FUGG. C.D. gratefully acknowledges the support of the Bavarian Academy of Sciences and Humanities.
\end{acknowledgments}

\bibliography{Papers}

\end{document}